\documentclass[12pt]{article}

\usepackage{amssymb}
\usepackage{amsmath}
\usepackage{amsfonts}

\newcommand{\R}{\mathbb{R}}
\newcommand{\C}{\mathbb{C}}

\newcommand{\N}{\mathbb{N}}

\newcommand{\be}{\begin{equation}}
\newcommand{\ee}{\end{equation}}
\newcommand{\bea}{\begin{eqnarray}}
\newcommand{\eea}{\end{eqnarray}}
\newcommand{\kt}{\rangle}
\newcommand{\br}{\langle}

\begin{document}

\title{${\cal PT}$-Symmetric
Quantum Mechanics: A Precise and Consistent Formulation}
\author{\\
Ali Mostafazadeh\thanks{E-mail address: amostafazadeh@ku.edu.tr}\\
\\ Department of Mathematics, Ko\c{c} University,\\
Rumelifeneri Yolu, 34450 Sariyer,\\
Istanbul, Turkey}
\date{ }
\maketitle

\begin{abstract}
The physical condition that the expectation values of physical
observables are real quantities is used to give a precise
formulation of ${\cal PT}$-symmetric quantum mechanics. A
mathematically rigorous proof is given to establish the physical
equivalence of ${\cal PT}$-symmetric and conventional quantum
mechanics. The results reported in this paper apply to arbitrary
${\cal PT}$-symmetric Hamiltonians with a real and discrete
spectrum. They hold regardless of whether the boundary conditions
defining the spectrum of the Hamiltonian are given on the real
line or a complex contour.
\end{abstract}

\section{Introduction}

Perhaps one of the most important lessons I have learned during my
mathematical education, and highly appreciated in my scientific
carrier, is that ``{\em The most difficult statements to prove are
those that are false}, \cite{ulger}.'' In my study of theoretical
physics I also learned that a prominent feature of all successful
fundamental theories is that they have a rigid structure. This
actually constitutes the basis of one of the main arguments for
justifying research in string theory.\footnote{There are a handful
of string theories that are known not to be inconsistent;
slightest variation of these leads to the violation of one or more
of the basic physical principles such as unitarity or absence of
anomalies.} What is not so well-known or at least
well-appreciated, especially among the contributors to the
development of ${\cal PT}$-symmetric QM, is that the conventional
QM itself is one of the most rigid physical theories that we have
come to establish. This is actually at the heart of the failure of
three generations of theoretical physicists to generalize QM in
any fundamental way. The purpose of the present paper is to show
that ${\cal PT}$-symmetric or complex extension of QM \cite{bbj}
is no exception.

As this has been a sensitive issue, I have tried to make my
analysis as precise and mathematically rigorous as possible. In
order to make the paper reasonably self-contained and avoid
potential terminological ambiguities, I have included the relevant
mathematical definitions and theorems. These should be easily
accessible for a typical theoretical physicist familiar with basic
features of QM.

The main motivation for the present study has been my attempts to
apply the following principle to ${\cal PT}$-symmetric QM: {\em In
order to achieve a satisfactory understanding of the virtues of a
new theoretical scheme, especially if it lacks experimental
support, one is obliged to formulate and state the basic
postulates of the theory in a precise language, translate them
into mathematical statements, and learn and use the standard
mathematical notions and theorems, or develop new mathematical
tools if necessary, to assess its viability as a consistent
physical theory and to determine its relation to the established
theories.}

I close this section by a quote from Jurg Fr\"ohlich: ``{\em It is
possible to do good theoretical physics and at the same time be
mathematically rigorous.}''\footnote{Stated during his lecture at
Les Houches Summer School on `Quantum Field Theory: Perspectives
and Prospective,' June 1998.}

\section{Basic Mathematical Facts}

The paragraphs of this section have been labelled for future
reference.

P0.~\textbf{Notations and Conventions:} $\N$, $\R$, and $\C$
denote the sets of natural (including zero), real, and complex
numbers, respectively; $\R^+$ stands for the set of strictly
positive real numbers; The symbols $\forall$, $\exists$, $^*$
respectively mean `for all', `there exists', and `complex
conjugate'; $\delta_{mn}$ stands for the Kronecker delta function.

P1.~Let $V$ be a complex vector space \cite{axler}. A function
$\br\cdot,\cdot\kt:V\times V\to\C$ is called an \textbf{inner
product} on $V$ if it is nondegenerate ($\br\phi,\psi\kt=0$ for
all $\phi\in V$ implies $\psi=0$), Hermitian
($\br\psi,\phi\kt=\br\phi,\psi\kt^*$), and sesquilinear ($\forall
a,b\in\C,~\forall\psi,\phi,\chi\in
V,~\br\psi,a\phi+b\chi\kt=a\br\psi,\phi\kt+b\br\psi,\chi\kt$). An
inner product $\br\cdot,\cdot\kt$ is called
\textbf{positive-definite} if it also satisfies:
$\br\psi,\psi\kt\in\R^+\cup\{0\}$ for all $\psi\in V$, and
$\br\psi,\psi\kt=0$ implies $\psi=0$.\footnote{As this condition
implies nondegenerateness, usually one does not separately
postulate the latter in defining a positive-definite inner
product.}

P2.~A \textbf{complex inner product space} $N$ is a complex vector
space endowed with a positive-definite inner product. The function
$\parallel\cdot\parallel: V\to\R^+\cup\{0\}$ defined by
$\parallel\psi\parallel:=\sqrt{\br\psi,\psi\kt}$ is called the
\textbf{norm} of $N$. The \textbf{convergence} of sequences
$\{\psi_n\}$ in $N$ is defined using the norm: $\psi_n\to\psi$ if
$\lim_{n\to\infty}\parallel\psi_n-\psi\parallel=0$. $\{\psi_n\}$
is called a \textbf{Cauchy sequence} if $\forall m,n\geq M$,
$\lim_{M\to\infty}\parallel\psi_n-\psi_m\parallel=0$ (i.e.,
$\forall\epsilon\in\R^+,~\exists M\in\N$ such that $\forall
m,n\geq M, \parallel\psi_n-\psi_m\parallel<\epsilon$.) $A\subseteq
N$ is said to be a \textbf{dense subset} of $N$ if every $\psi\in
N$ is the limit of a sequence of elements of $A$.

P3.~A complex \textbf{Hilbert space} ${\cal H}$ is a complex inner
product space that is norm-complete, i.e., the Cauchy sequences in
${\cal H}$ converge. All the Hilbert spaces used in this paper are
complex Hilbert spaces.

P4.~There is a well-defined procedure to extend an inner product
space $N$, in a unique way, to a Hilbert space ${\cal H}$ such
that $N$ is dense in ${\cal H}$, \cite{reed-simon}. ${\cal H}$ is
then called the \textbf{Cauchy-completion} of $N$, and $N$ is said
to be Cauchy-completed to ${\cal H}$.

P5.~A complex \textbf{separable Hilbert space} is a complex
Hilbert space that has a countable orthonormal basis. The latter
is a subset ${\cal B}=\{\beta_n\in{\cal H}|n\in\N\}$ of ${\cal H}$
satisfying: $\forall m,n\in\N$,
$\br\beta_m,\beta_n\kt=\delta_{mn}$ and $\forall\psi\in{\cal H}$,
$\exists a_n\in\C$ such that $\psi=\sum_{n=0}^\infty a_n \beta_n$.
The typical examples are the space of square-integrable functions:
$L^2(\R):=\{\psi:\R\to\C|\int_{\R}|\psi(x)|^2dx<\infty\}$ with
inner product $\br\psi|\phi\kt=\int_{\R}\psi(x)^*\phi(x)dx$, and
the space of square-summable sequences:
$l_2:=\{\psi:\N\to\C|\sum_{n=0}^\infty|\psi(n)|^2<\infty\}$ with
inner product $(\psi,\phi)=\sum_{n=0}^\infty\psi(n)^*\phi(n)$.

P6.~Let ${\cal H}_i$, with $i\in\{1,2\}$, be inner product spaces
having inner products $\br\cdot,\cdot\kt_i$. Then the
\textbf{adjoint} $L^\dagger:{\cal H}_2\to{\cal H}_1$ of a linear
operator $L:{\cal H}_1\to{\cal H}_2$ is the unique linear map
satisfying: $\br\psi_2,L\psi_1\kt_2=\br
L^\dagger\psi_2,\psi_1\kt_1$, $\forall\psi_i\in{\cal H}_i$. $L$ is
called \textbf{unitary} if (C1:) $\forall\psi,\phi\in{\cal H}_1$
one has $\br L\psi,L\phi\kt_2=\br\psi,\phi\kt_1$,
\cite{reed-simon}. C1 is equivalent to (C2:) $L^\dagger L=I_1$,
where $I_1$ is the identity operator for ${\cal H}_1$.\footnote{I
was surprised to see the author of \cite{wang} referred to C1 as a
generalization of the standard definition of unitarity. I was
amazed to read a referee report on \cite{critique} saying that ``a
huge majority of mathematical physicists'' will not accept C1
because it did not imply C2, and that I was ``WRONG'' not to warn
the reader of the difference and to use C2 in another publication
that I cited in \cite{critique}!} Another equivalent condition is
(C3:) $L$ maps elements of any orthonormal basis of ${\cal H}_1$
to those of an orthonormal basis of ${\cal H}_2$ in a one-to-one
and onto manner. A linear operator $K:{\cal H}_1\to{\cal H}_1$ is
called \textbf{self-adjoint} or \textbf{Hermitian} if
$K^\dagger=K$, i.e., $K$ satisfies $\br\psi,K\phi\kt_1=\br
K\psi,\phi\kt_1$, $\forall\psi,\phi\in{\cal H}_1$.\footnote{In
most mathematical texts, e.g., \cite{reed-simon}, what we call a
Hermitian operator is referred to as a `symmetric operator'. In
${\cal PT}$-symmetric QM as described in \cite{bbj,bbj-ajp} a
symmetric operator means an operator that is represented by a
symmetric matrix in the usual position representation.}

P7.~Two inner product (in particular Hilbert) spaces ${\cal
H}_1,{\cal H}_2$ are said to be \textbf{unitarily equivalent} if
there is a unitary operator $U:{\cal H}_1\to{\cal H}_2$. In this
case a linear operator $K_1:{\cal H}_1\to{\cal H}_1$ is Hermitian
if and only if $K_2:=UK_1U^{-1}$ is a Hermitian operator acting in
${\cal H}_2$.

P8.~Let ${\cal B}=\{\beta_n\}$ be a basis of a separable Hilbert
space ${\cal H}$ and $K:{\cal H}\to{\cal H}$ be a linear map. The
infinite matrix $K^{({\cal B})}$ with entries $K^{({\cal
B})}_{mn}$ defined by $K\beta_n=\sum_{m=0}^\infty K^{({\cal
B})}_{mn} \beta_m$ is called the \textbf{matrix representation} of
$K$ in the basis ${\cal B}$. Now, suppose that ${\cal B}$ is
orthonormal, in which case $K^{({\cal B})}_{mn}=
\br\beta_m,K\beta_n\kt$. Then, $K$ is Hermitian if and only if
$K^{({\cal B})}$ is a Hermitian matrix, i.e., $K^{({\cal
B})}_{nm}={K^{({\cal B})}_{mn}}^*$. \textbf{For a non-orthonormal
basis ${\cal B}$, there is no relationship between the Hermiticity
of $K$ and the Hermiticity of $K^{({\cal B})}$.}

P9.~The condition that a given basis ${\cal B}$ of a separable
Hilbert space ${\cal H}$ be orthonormal fixes the inner product on
${\cal H}$ uniquely. This together with the fact that every
separable Hilbert space has an orthonormal basis lead to the
\textbf{uniqueness theorem for separable Hilbert spaces}
\cite{reed-simon}: {\em Up to unitary equivalence there is a
unique separable Hilbert space}, i.e., any pair of separable
Hilbert spaces ${\cal H}_1$ and ${\cal H}_2$ are related by a
unitary operator $U:{\cal H}_1\to{\cal H}_2$.

\section{${\cal PT}$-Symmetric QM as a Fundamental Theory}
\subsection{Bender's Formulation}

Consider the ${\cal PT}$-symmetric Hamiltonians of the form
$H=p^2+v(ix)$ where $v:\R\to\R$ is a potential. A typical example
is $v(x)=-\mu x^2-\lambda x^{2+\epsilon}$, with
$\mu,\lambda,\epsilon\in\R$, which for $\mu=0$, $\lambda=1$ yields
\cite{bbj}
    \be
    H=p^2+x^2(ix)^\epsilon.
    \label{H}
    \ee
There is a certain class of potentials $v$ for which the
corresponding Sturm-Liouville problem for the Hamiltonian,
$-\phi''(x)+v(ix)\phi(x)=E\phi(x)$, may be shown to lead to a
strictly real, positive, discrete and nondegenerate spectrum,
provided that the vanishing boundary conditions at infinity are
imposed along an appropriate contour $C$ in the complex plane,
\cite{dorey-shin}. For the Hamiltonian (\ref{H}), this happens
whenever $\epsilon\geq 0$, \cite{bender2}.

Suppose that $v$ belongs to this class and $C$ is chosen so that
the corresponding spectrum for $H$ is strictly real, discrete, and
nondegenerate.\footnote{For physical reasons, one should also
require that the spectrum be bounded from below.} Let ${\cal
B}:=\{\phi_n\}$ be a set of ${\cal PT}$-invariant \cite{bbj}
eigenfunctions of $H$ with distinct eigenvalues $E_n$.

The claim upon which the ${\cal PT}$-symmetric QM rests is that
one can define a Hilbert space for a quantum system whose
dynamics, as determined by the Schr\"odinger equation:
$i\frac{d\psi(t)}{dt}=H\psi(t)$, is unitary \cite{bbj,bbj-ajp}.
$H$ serves as the quantum Hamiltonian operator for this system. In
particular, the states of the system are described by
superpositions of $\phi_n$, and its energy levels have energies
$E_n$. This implies that, as a (complex) vector space, the space
$V$ of state vectors is the span of ${\cal B}$. In \cite{bbj}, it
is shown that one can define a positive-definite (${\cal CPT}$-)
inner product $\br\cdot,\cdot\kt_+$ on $V$, in such a way that
${\cal B}$ is promoted to an orthonormal basis. By construction,
this defines an infinite-dimensional complex inner product space
$N$ having a countable basis. It is this space that Bender and his
coworkers \cite{bbj} identify with the `physical Hilbert space' of
the system.

The physical observables were initially defined by Bender et al
\cite{bbj,bbj-ajp,bbj-jpa} as ${\cal CPT}$-invariant operators,
i.e., those that commute with ${\cal CPT}$, (Def.~1). In
\cite{critique}, I showed that this definition was inconsistent
with the dynamics of the theory and proposed to identify the
observables with Hermitian operators acting in the physical
Hilbert space (Def.~2). In their recent Erratum \cite{bbj-erratum}
to \cite{bbj}, Bender et al give an alternative definition, namely
(Def.~3:) observables are linear operator $A$ satisfying
$A^T={\cal CPT} A\,{\cal CPT}$, where transposition $^T$ is
defined in the usual position representation according to
$A^T(x,x')=A(x',x)$.\footnote{The position representation
$O(x,x')$ of a linear operator $O$ is defined by
$(O\psi)(x):=\int_\R O(x,x')\psi(x')dx'$.} In \cite{comment}, I
discuss the shortcomings of Def.~3 and show that even if they
could be resolved for a given system then Def.~3 reduces to a
special case of Def.~2.

Note that Def.~1 formed the basis of the rather appealing idea
that in ${\cal PT}$-symmetric QM one could ``replace the
mathematical condition of Hermiticity, whose physical content is
somewhat remote and obscure, by the physical condition of
space-time and charge-conjugation symmetry, \cite{bbj,bbj-ajp}.''
Def.2 clearly defies this claim. The same is true for Def.~3,
because it requires the Hamiltonian to be symmetric, a property as
mathematical/non-physical as being Hermitian. In fact, Def.~3
makes explicit use of the operation of transposition $^T$ which,
according to \cite{bbj,bbj-ajp}, is related to Hermitian
conjugation $^\dagger$ as $A^\dagger={\cal T}\,A^T{\cal T}$.

\subsection{A Complete and Consistent Formulation}

To make the above statement that `$N$ is the physical Hilbert
space of the system' meaningful, one is forced to Cauchy complete
$N$ to a Hilbert space ${\cal H}$. There is essentially no other
mathematically viable alternative that would be consistent with
the physical principle that the state vectors $\phi$ are
superpositions of (possibly infinite number of) energy
eigenfunctions $\phi_n$, i.e., $\phi=\sum_{n=0}^\infty a_n\phi_n$
for some $a_n\in\C$. As $N$ is spanned by ${\cal B}$ and is dense
in ${\cal H}$, ${\cal B}$ is necessarily a complete orthonormal
basis of ${\cal H}$. It is also a countable set. Therefore,
according to P5, ${\cal H}$ is a separable Hilbert space. This in
turn implies, in view of P9, that ${\cal H}$ is unitarily
equivalent to both $L^2(\R)$ and $l_2$. As I show below it is easy
to construct the unitary operators realizing these equivalences.

To establish the unitary equivalence of ${\cal H}$ and $l_2$, I
use the basis ${\cal B}$. Let $U:{\cal H}\to l_2$ be defined by
$(U\psi)(n):=\br\phi_n,\psi\kt_+$, $\forall n\in\N$. A simple
calculation shows that $\forall\psi,\phi\in{\cal H}$,
$(U\psi,U\phi)=\sum_{n=0}^\infty
\br\phi_n,\psi\kt_+^*\br\phi_n,\phi\kt_+=\br\psi,\phi\kt_+$, where
I have used the completeness of ${\cal B}$. This proves that $U$
satisfies condition C1 (and hence C2) of P6; it is a unitary
operator.

Let $e_n:= U\phi_n$, then $e_n(m)=\br\phi_m,\phi_n\kt_+=
\delta_{mn}$. This shows that $U$ maps the basis ${\cal B}$ onto
the standard orthonormal basis $\{ e_n\}$ of $l_2$. Furthermore,
$U$ maps the Hamiltonian $H$ to the linear operator $UHU^{-1}$
that acts in $l_2$. This is just the matrix representation of $H$
in the basis ${\cal B}$, $UHU^{-1}=H^{({\cal B})}$, with entries
$H^{({\cal B})}_{mn}= \br\phi_m,H\phi_n\kt=E_n\delta_{mn}$.
Because $E_n\in\R$, $H^{({\cal B})}$ is a Hermitian matrix. This
together with the fact that ${\cal B}$ is orthonormal imply that
\textbf{$H$ is a Hermitian operator acting in ${\cal H}$}. This is
clearly consistent with Def.~2 of a physical observable; $H$ is an
observable.

In \cite{critique}, I show that Def.~2 is the only definition that
is consistent with the requirements of the quantum measurement
theory. Here I give an alternative and shorter proof of this
statement. It uses the following well-known theorem of linear
algebra. {\bf Theorem:} {\em Let $H:N\to N$ be a linear operator
acting in an inner product space $N$ with inner product
$\br\cdot,\cdot\kt$. Then $H$ is Hermitian if and only if for all
$\psi\in N$, $\br\psi,H\psi\kt$ is real},  \cite{axler}. This
theorem together with the physical postulate that the expectation
values of the observables must be real numbers identify Def.~2 as
the only physically acceptable definition of an observable. In the
rest of this paper I use Def.~2 for a physical observable.

The unitary operator $U$ may be used to obtain all the observables
of the theory in the original ${\cal PT}$-symmetric picture,
\cite{critique}. According to P7, the observables $O$ acting in
the Hilbert space ${\cal H}$ are given by $O
\psi:=\sum_{n=0}^\infty O_{mn}\br\phi_m,\psi\kt_{_+}\,\phi_n$,
where $O_{mn}=O_{nm}^*\in\C$, i.e., $O$ has a Hermitian matrix
representation in the basis ${\cal B}$. Clearly, this is
consistent with P8.

As explained in \cite{critique} the statement that the Hamiltonian
(\ref{H}) is non-Hermitian stems from the definition used by
Bender and his collaborators which is equivalent to saying that
$H$ is not a Hermitian operator as an operator acting in
$L^2(\R)$. Because $L^2(\R)$ is not the physical Hilbert space for
the problem, this statement does not have any physical
significance. However, it is also not true that one cannot
describe the same physical system using conventional quantum
mechanics based on the Hilbert space $L^2(\R)$.

Let ${\cal U}:{\cal H}\to L^2(\R)$ be the map ${\cal U}\psi:=
\sum_{n=0}^\infty \br\phi_n,\psi\kt_+ \Phi_n$ where ${\cal
F}:=\{\Phi_n\}$ is a fixed complete orthonormal basis of
$L^2(\R)$, e.g., one can identify $\Phi_n$ with the normalized
eigenfunctions of the simple harmonic oscillator Hamiltonian with
unit mass and frequency (in some appropriate units). One can
easily show, using the condition C3 of P6, that ${\cal U}$ is a
unitary operator, for by construction it maps $\phi_n$ to $\Phi_n$
and both ${\cal B}$ and ${\cal F}$ are complete orthonormal bases.
A simple calculation also shows that under the action of ${\cal
U}$, the Hamiltonian $H$ maps to
    \be
    h:={\cal U}H{\cal U}^{-1}=\sum_{n=1}^\infty
    E_n|\Phi_n\kt\br\Phi_n|,
    \label{hermitian-h}
    \ee
where I have used the standard bra-ket notation in $L^2(\R)$.
Clearly $h$ is a Hermitian operator acting in $L^2(\R)$.
Similarly, the observables $O$ are related to Hermitian operators
$o:L^2(\R)\to L^2(\R)$ according to
    \be
    O={\cal U}^{-1}o\:{\cal U}.
    \label{hermitian-o}
    \ee
So that $O_{mn}:=\br\phi_m,O\phi_n\kt_+=\br\Phi_m|o|\Phi_n\kt$.

This completes the formulation of the ${\cal PT}$-symmetric
quantum mechanics: The Hilbert space, the Hamiltonian, and the
observables are determined, and the physical interpretation is
provided by the standard measurement theory.

\subsection{Equivalence with Conventional QM}

Let $H$, ${\cal H}$, $O$, and ${\cal U}$ be as in the preceding
subsection. Consider computing the expectation value ${\cal O}(t)$
of an observable $O$ in the state described by $\psi(t)\in {\cal
H}$ for a given initial state vector $\psi(t_0)\in {\cal H}$. The
result is
    \be
    {\cal O}(t)=\frac{\br\psi(t_0)e^{i(t-t_0)H},O
    e^{-i(t-t_0)H}\psi(t_0)\kt_+}{
    \br\psi(t_0),\psi(t_0)\kt_+},
    \label{exp-1}
    \ee
where I have used the fact that the time-evolution operator
$e^{-i(t-t_0)H}$ is a unitary operator acting in ${\cal H}$.

Next, let $\Psi(t_0):={\cal U}\psi(t_0)$ and use the unitary
operator ${\cal U}$ to compute the same quantity. In view of
(\ref{hermitian-h}), (\ref{hermitian-o}) and C1 of P6 we have,
    \be
    {\cal O}(t)=\frac{\br\Psi(t_0)|e^{i(t-t_0)h}\,o\:
    e^{-i(t-t_0)h}|\Psi(t_0)\kt}{
    \br\Psi(t_0)|\Psi(t_0)\kt}.
    \label{exp-2}
    \ee
Eqs.~(\ref{exp-1}) and (\ref{exp-2}) show that the ${\cal
PT}$-symmetric description of the physical system that uses
$({\cal H},H,O)$ is physically equivalent to the one that uses
$(L^2(\R),h,o)$. This establishes the equivalence of ${\cal
PT}$-symmetric and conventional QM as fundamental physical
theories. This equivalence is a manifestation of the rigidity of
the basic structure of QM that is referred to in Sec.~1.

The choice between the above two equivalent representations in the
description of a physical system is subjective in nature. One
might argue that the ${\cal PT}$-symmetric description is more
advantageous, for the Hamiltonian $H$ has a much simpler form.
This is true if one is interested in finding the energy levels of
the system. However, in general, for example in computing the
expectation value ${\cal O}(t)$, the two representations involve
the same level of practical difficulties. Though it might be
easier to compute the time-evolution operator for $H$ in
(\ref{exp-1}), the computation of $O$ (from $o$) and the
evaluation of the inner products appearing in (\ref{exp-1}) are as
difficult as the computation of $o$ (from $O$) and $h$ that appear
in (\ref{exp-2}).

\section{Conclusion}

I have provided a complete formulation of ${\cal PT}$-symmetric QM
and gave a rigorous proof that, as a fundamental physical theory,
it is equivalent to the conventional QM. I had previously used the
machinery of pseudo-Hermitian Hamiltonians to establish the
existence of a unitarily equivalent Hermitian Hamiltonian
associated with a given ${\cal PT}$-symmetric Hamiltonian,
\cite{jpa-2003}. The construction of the Hilbert space ${\cal H}$
and the fact that the Hamiltonian is a Hermitian operator acting
in ${\cal H}$ have also been discussed in \cite{kresh}. The
results of \cite{jpa-2003,kresh} show the equivalence of the
dynamical structure of the ${\cal PT}$-symmetric and conventional
QM. In the present paper I have given an explicit proof
establishing the equivalence of the dynamical as well as the
kinematical structures of the two theories. Note that this proof
does not rely on the notion of pseudo-Hermiticity.

The results reported in this paper clearly show that one should
not expect to get more from the ${\cal PT}$-symmetric QM than what
conventional QM has to offer. This may sound as a negative and
discouraging statement, especially for those (like me) who have
invested a great deal of time and effort to understand ${\cal
PT}$-symmetric QM and helped develop it further. Personally, I
would prefer to view the matter from a different angle: The above
equivalence also means that ${\cal PT}$-symmetric QM is as
valuable as the conventional QM. Moreover, this equivalence holds
only if one insists in viewing ${\cal PT}$-symmetric QM as a
fundamental theory. I consider the interesting mathematical
content of ${\cal PT}$-symmetric QM as sufficient evidence that it
may find good uses in the study of various effective theories. It
is also important to note that the pseudo-Hermitian operators
\cite{p1}, that were primarily developed to deal with ${\cal
PT}$-symmetric QM, have already found remarkable applications in
relativistic QM \cite{cqg,rqm} and quantum cosmology
\cite{cqg,ap}. A more recent related development has been a
formulation of a unitary QM based on a time-dependent Hilbert
space \cite{pla-2004}. This has led to a new interpretation for
geometric phases and revealed an interesting similarity between QM
and General Relativity. To these I should also add the
applications reported in \cite{ahmed}. Finally, I must emphasize
that the results of this paper apply to ${\cal PT}$-symmetric QM.
The status of ${\cal PT}$-symmetric QFT \cite{bbj,bbj-ajp,b} is
completely open to future investigations.

\section*{Acknowledgment}

This work has been supported by the Turkish Academy of Sciences in
the framework of the Young Researcher Award Program
(EA-T$\ddot{\rm U}$BA-GEB$\dot{\rm I}$P/2001-1-1).

\newpage

{\small

\end{document}